\documentclass[aps,twocolumn,groupedaddress,showpacs]{revtex4-1}
\usepackage{graphics}
\usepackage{epstopdf}
\usepackage{soul}
\usepackage{sfmath}
\usepackage{amsmath}
\begin{document}
\renewcommand{\familydefault}{cmss}
\sffamily
\newcommand{\op}{\boldsymbol}
\newcommand{\beq}{\begin{equation}}
\newcommand{\eeq}{\end{equation}}
\bibliographystyle{apsrev}

\title{Quantum Eraser for Three-Slit Interference}
\author{Naveed Ahmad Shah}
\email{shahnaveed75@gmail.com}
\affiliation{Department of Physics, Jamia Millia Islamia, New Delhi-110025, India.}
\author{Tabish Qureshi}
\email{tabish@ctp-jamia.res.in}
\affiliation{Centre for Theoretical Physics, Jamia Millia Islamia, New Delhi-110025, India.}

\begin{abstract}

It is well known that in a two-slit interference experiment, if
the information, on which of the two paths the particle followed, is
stored in a quantum path detector, the interference is destroyed. However,
in a setup where this path information is ``erased'', the interference
can reappear. Such a setup is known as a quantum eraser. A generalization of
quantum
eraser to a {\em three-slit} interference is theoretically analyzed.
It is shown that three complementary interference patterns can arise out of the quantum
erasing process. 


\end{abstract}
\maketitle

\section{Introduction}

The first double-slit interference experiment was performed with light by Thomas Young
in 1801, thereby demonstrating the wave nature of light \cite{young}. With
the advent of
quantum theory, it was realized that all quantum particles, although
considered indivisible, should show wave nature. The first double-slit
interference with electrons was demonstrated by J\"onsson \cite{jonsson}.

A double-slit experiment with massive quantum particles brings in a whole new concept,
that of a particle interfering with itself. That the interference of
electrons in a double-slit experiment involves an electron interfering
with itself, and not with other electrons, was conclusively demonstrated
for the first time by Tonomura et.al. \cite{tonomura}.

Niels Bohr had stated that the wave aspect and the particle aspect are
complementary, in the sense that if an experiment clearly reveals the wave
nature, it will completely hide the particle aspect and vice-versa \cite{bohr}.
The complementarity principle came under attack right after
its inception when Einstein proposed his famous recoiling slit experiment
(see e.g. \cite{tqeinstein}). Since then it has been a subject of debate
regarding it's various aspects and it's validity too \cite{dsen,wootters,greenberger}.
The current
understanding is that in the context of the two-path interference experiment,
either as a two-slit experiment, or as a Mach-Zhender interferometer,
the principle of complementarity is quantitatively represented by the
so-called duality relation \cite{englert}
\begin{equation}
{\mathcal V}^2 + {\mathcal D}^2 \le 1,
\label{duality}
\end{equation}
where  ${\mathcal D}$ is a path distinguishability and ${\mathcal V}$ the
visibility of the interference pattern. 

A very interesting consequence of this wave-particle duality is the 
so-called quantum eraser \cite{jaynes,scully}. The idea of quantum
eraser is that if the which way information can, in some way, be ``erased'',
the lost interference pattern can be made to reappear. This phenomenon holds
even when the which-path information has been erased {\em well after} 
the particle has been registered on the screen. In such a scenario,
the phenomenon is called ``delayed-choice'' quantum eraser. In course of
time, quantum eraser was experimentally demonstrated for two-path
interference in several different ways \cite{eraser1,eraser2,eraser3,eraser4,eraser5,eraser6,eraser7,eraser8,eraser9,eraser10,eraser11,eraser12}

\begin{figure}
\centerline{\resizebox{9cm}{!}{\includegraphics{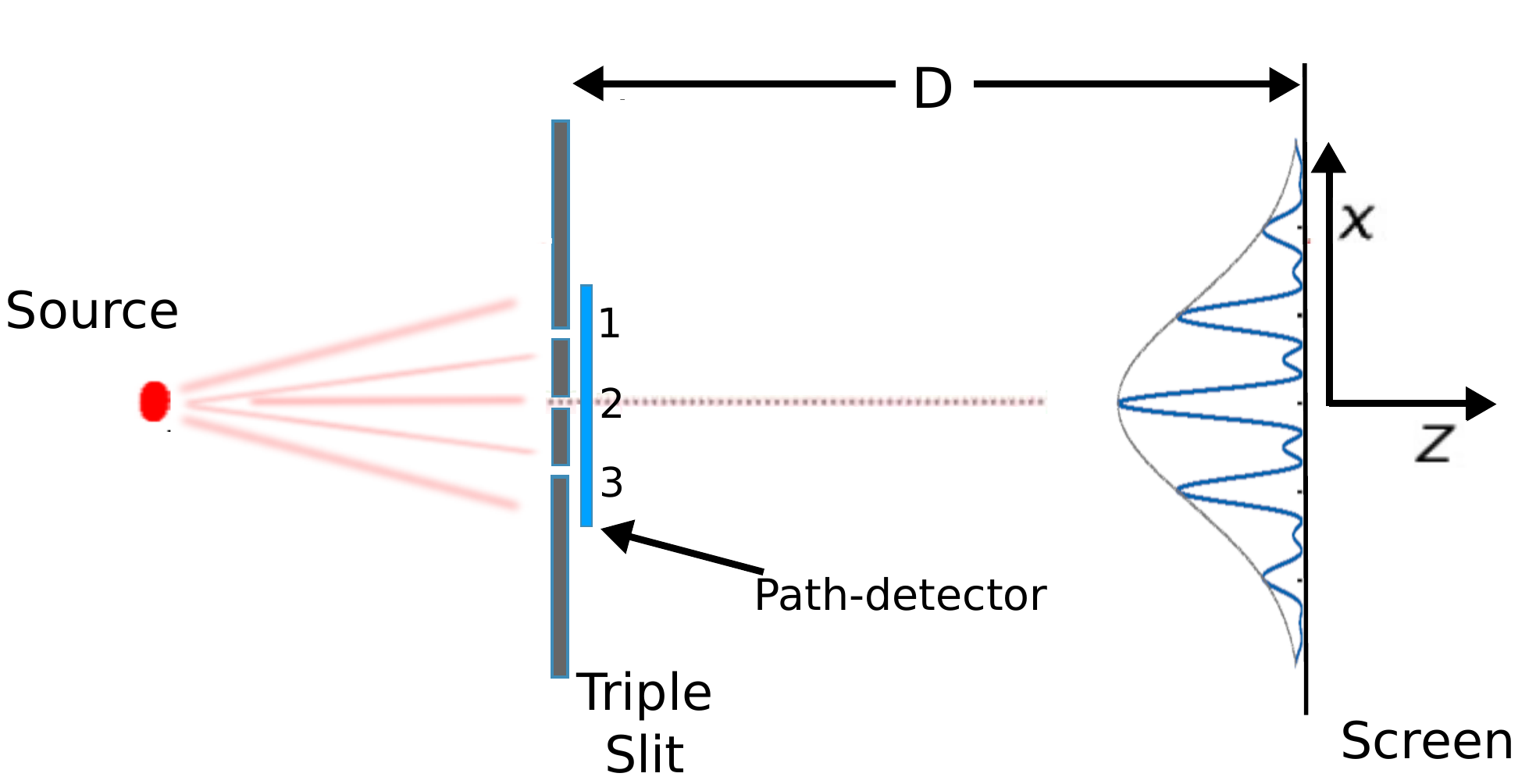}}}
\caption{Schematic diagram of a triple-slit interference experiment. 
A quantum path-detector could be added to the setup, which is capable
of obtaining information on which slit the particle passed through.}
\label{tslit}
\end{figure}

The concept of complementarity should hold in multi-path or multi-slit 
experiments too. It is an issue which has been, and continues to be,
a focus of much research attention 
\cite{jaeger,durr,bimonte,englertmb,3slit,g3slit,coherence,bagan,nslit}.
We pose the question, is quantum eraser possible in multislit interference
experiments too, and are there any subtleties involved? We begin by
looking at the issue of quantum erasing in a 3-slit interference experiment,
which is the subject of this investigation.

\section{Three-slit interference and wave-particle duality}

Three slit interference is a bit more complex than its two-slit counterpart,
because it involves superposition of three parts (see Fig. \ref{tslit}).
A particle emerging from a triple-slit may be assumed to have a form
\begin{equation}
|\psi\rangle = {1\over \sqrt 3}|\psi_1\rangle + {1\over \sqrt 3}|\psi_2\rangle
+ {1\over\sqrt{3}}|\psi_3\rangle
\label{pure0}
\end{equation}
where $|\psi_1\rangle, |\psi_2\rangle, |\psi_3\rangle$ represent the state
of the particle corresponding to it coming out of slit 1, 2 and 3, respectively.
Sharpest interference is obtained when the amplitudes for the three 
possibilities are equal, which is $1/\sqrt{3}$ in our case. By virtue of the
Born rule, the three-slit interference can be thought of as arising from
three two-slit interferences \cite{urbasi}.

If one were to introduce a {\em quantum} path-detector which can tell which
of the three slits the particle has gone through, the state of the particle
and the path-detector may be written as
\begin{equation}
|\psi\rangle = {1\over \sqrt 3}|\psi_1\rangle|+\rangle + {1\over \sqrt 3}|\psi_2\rangle|0\rangle
+ {1\over\sqrt{3}}|\psi_3\rangle|-\rangle
\label{ent0}
\end{equation}
where $|+\rangle, |0\rangle, |-\rangle$ represent three orthonormal states
of the path-detector. If the path-detector is capable of gaining information
on which slit the particle went through, by virtue of von Neumann's idea
of a quantum measurement \cite{neumann}, the combined state has to have the
above form.

If the state of a particle, emerging from the triple slit, is given by
(\ref{pure0}), the state of the particle at a time $t$ when it reaches the
screen, would be different because it would have undergone a time-evolution.
Using quantum wave-packets is a useful way to study the dynamics of
quantum particles (see e.g. \cite{kaur}).
In order do an accurate analysis of the interference, we carry out a 
wave-packet analysis of the particle. We assume that the states emerging
from the three slits are Gaussian wave-packets, centered at the respective
slits, with a width which is related to the width of the slits.
Assuming that the particle travels along the z-axis, and the three slits are
parallel to the y-axis, located at $x=d$, $x=0$, and $x=-d$.
In order to keep the analysis simple, we ignore the dynamics along the
z- and y-axes. The dynamics along the z-axis serves only to transport the
particle from the slits to the screen. We assume that the particle travels
in the z-direction with an average momentum $p_0$, and a deBroglie wavelength
$\lambda = h/p_0$ is associated with it. The particle travels a distance
$D$ in a time $t$, and $D = p_0t/m$.  This leads us to
\begin{eqnarray}
 \lambda D = ht/m.
\end{eqnarray}
The more interesting dynamics is in the
x-direction where the wave-packets are expected to expand and overlap,
giving rise to interference.  One can go beyond this approximation by
treating the dynamics along the y- and z-direction more rigorously \cite{beau}.
However, for our purpose, this approximate treatment suffices.

The states of the particle in front of the three slits, just after it
emerges, are assumed to be
\begin{eqnarray}
\langle x|\psi_1\rangle &=& C \exp\left({-(x-d)^2\over \epsilon^2}\right)\nonumber\\
\langle x|\psi_2\rangle &=& C \exp\left({-x^2\over \epsilon^2}\right)\nonumber\\
\langle x|\psi_3\rangle &=& C \exp\left({-(x+d)^2\over \epsilon^2}\right),
\end{eqnarray}
where $C = (2/\pi\epsilon^2)^{1/4}$. The state (\ref{pure0}) can then be
represented as
\begin{equation}
\langle x|\psi\rangle = {C\over \sqrt 3}\left(e^{{-(x-d)^2/ \epsilon^2}}
+ e^{{-x^2/ \epsilon^2}}
+ e^{{-(x+d)^2/ \epsilon^2}} \right).
\label{purex0}
\end{equation}
We assume the particle travels freely, and has a mass $m$, and hence it's
evolution is governed by the operator
\begin{equation}
\op{U}(t) = \exp(-i\op{p}^2t/2m\hbar),
\end{equation}
where $\op{p}$ represent the operator for the x-component of the momentum
of the particle. Using the time-evolution governed by the above, the 
state of the particle, when it reaches the screen after a time $t$, is
given by
\begin{eqnarray}
\psi(x,t) = {C_t\over \sqrt 3}&&\left(e^{{-(x-d)^2/(\epsilon^2+ia)}}
+ e^{{-x^2/(\epsilon^2+ia)}}\right.\nonumber\\
&&\left.+ e^{{-(x+d)^2/(\epsilon^2+ia)}} \right),
\label{purext}
\end{eqnarray}
where $a = 2\hbar t/m = \lambda D/\pi$ and $C_t = ({2\over\pi(\epsilon+ia/\epsilon)})^{1/4}$. 
The probability of the particle, hitting the screen at a position $x$, is
given by
\begin{eqnarray}
|\psi(x,t)|^2 &=& {|C_t|^2\over 3}e^{-2x^2/\Omega}\left(
1 + e^{-2d^2/\Omega}2\cosh(4xd/\Omega)\right.\nonumber\\
&&\left.+ e^{-(d^2-2xd)/\Omega}2\cos(2xd/a - d^2/a)  \right.\nonumber\\
&&\left.+ e^{-(d^2+2xd)/\Omega}2\cos(2xd/a + d^2/a)  \right.\nonumber\\
&&\left.+ e^{-2d^2/\Omega}2\cos(4xd/a) \right),
\label{purept}
\end{eqnarray}
where $\Omega = \epsilon^2+\lambda^2D^2/\pi^2\epsilon^2$, and is related 
to the expanded width of the wave-packets.
In the limit where the slits are very narrow and the wave-packets spread 
much much wider than $d$ and overlap with each other strongly, i.e.
$\Omega \gg d^2$, the above can be approximated by
\begin{eqnarray}
|\psi(x,t)|^2 &\approx& {|C_t|^2\over 3}e^{-2x^2/\Omega}\left(
1 + 2\cosh(4xd/\Omega)\right.\nonumber\\
&&\left.+ 4\cosh(2xd/\Omega)\cos(2xd/a)  \right.\nonumber\\
&&\left.+ 2\cos(4xd/a) \right)
\label{purepta}
\end{eqnarray}
The above expression represents a 3-slit interference pattern (see Fig. \ref{sxplot}).

In the presence of a path-detector, the state after the particle emerges
from the triple slit, is be given by
\begin{equation}
\langle x|\psi\rangle = {C\over \sqrt 3}\left(e^{{-(x-d)^2/ \epsilon^2}}
|+\rangle
+ e^{{-x^2/ \epsilon^2}}|0\rangle
+ e^{{-(x+d)^2/ \epsilon^2}}|-\rangle \right),
\label{entx0}
\end{equation}
which is just the position representation of (\ref{ent0}). After travelling
to the screen, the probability of the particle hitting it at a position $x$
is given by 
\begin{eqnarray}
|\psi(x,t)|^2 &=& {|C_t|^2\over 3}e^{-2x^2/\Omega}\left(
1 + e^{-2d^2/\Omega}2\cosh(4xd/\Omega)\right).\nonumber\\
\label{entpt}
\end{eqnarray}
One can see, that the cosine terms, which represented interference in
(\ref{purept}), are missing in the above expression. Basically they are
killed due to the orthogonality of $|+\rangle, |0\rangle, |-\rangle$.
The states $|+\rangle, |0\rangle, |-\rangle$ carry information about which slit
the particle passed through. So, it emerges that the storing of which-way
information is enough to destroy interference.
Wave-particle
duality in a 3-slit experiment has recently been quantitatively stated by a
new duality relation \cite{3slit}. However, here our focus is, looking at
the possibility
of quantum erasing in such an experiment, which is dealt with in the 
following discussion.

\section{Quantum eraser}

The three-state quantum path-detector may be thought of as a
pseudo-spin-1, whose z-component $\op{S_z}$ has the eigenstate
$|+\rangle, |0\rangle, |-\rangle$. These states can also be written in
terms of the eigenstates of the x-component of this pseudo-spin $\op{S_x}$,
which we represent as $|\uparrow\rangle, |\rightarrow\rangle, |\downarrow\rangle$. In this
representation, the eigenstates of $\op{S_z}$  look like
\begin{eqnarray}
|+\rangle &=& {1\over 2}|\uparrow\rangle +{1\over \sqrt{2}}|\rightarrow\rangle +{1\over 2}|\downarrow\rangle \nonumber\\
|0\rangle &=& {1\over \sqrt{2}}|\uparrow\rangle -{1\over \sqrt{2}}|\downarrow\rangle \nonumber\\
|-\rangle &=& {1\over 2}|\uparrow\rangle -{1\over \sqrt{2}}|\rightarrow\rangle +{1\over 2}|\downarrow\rangle
\end{eqnarray}
Using this, the state (\ref{ent0}) can be written as
\begin{eqnarray}
|\psi\rangle &=& {1\over \sqrt 3}\left[({|\psi_1\rangle\over 2}
+{|\psi_2\rangle\over \sqrt 2}+
{|\psi_3\rangle\over 2})|\uparrow\rangle
+ ({|\psi_1\rangle\over \sqrt 2}-{|\psi_3\rangle\over \sqrt 2})|\rightarrow\rangle
\right.\nonumber\\
&&\left. + ({|\psi_1\rangle\over 2} -{|\psi_2\rangle\over \sqrt 2}+
{|\psi_3\rangle\over 2})|\downarrow\rangle\right].
\label{entn0}
\end{eqnarray}

\begin{figure}
\centerline{\resizebox{10cm}{!}{\includegraphics{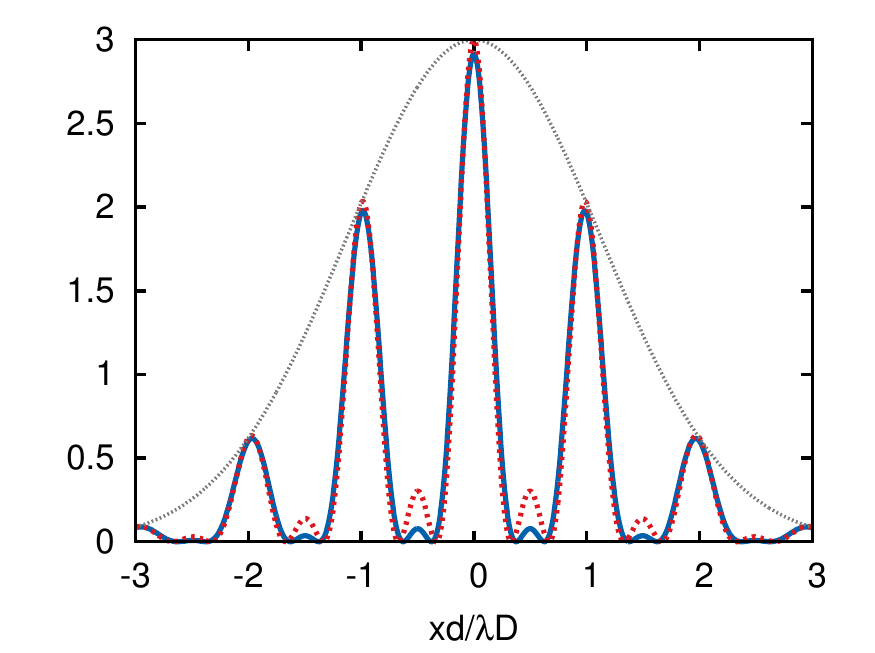}}}
\caption{Recovered interference pattern, given by $|\psi_{\uparrow}(x,t)|^2$
(solid line) and the original 3-slit interference pattern, given by
(\ref{purepta}) (dashed line).
The two are clearly different. The dotten line represents the lost
interference in the presence of which-way information, given by (\ref{entpt}).}
\label{sxplot}
\end{figure}

It is quite obvious that the above state will not lead to an interference
pattern, since it is the same as (\ref{ent0}). However, if one were to
measure the x-component of the pseudo-spin, and detect the particle in
correlation with the three results of the pseudo-spin measurement, the following
three situations will arise.
\begin{eqnarray}
\langle \uparrow|\psi(t)\rangle &=& {1\over \sqrt 3}\op{U}(t)\left({|\psi_1\rangle\over 2}
+{|\psi_2\rangle\over \sqrt 2}+ {|\psi_3\rangle\over 2}\right)\nonumber\\
\langle \downarrow|\psi(t)\rangle &=& {1\over \sqrt 3}\op{U}(t)\left({|\psi_1\rangle\over 2}
-{|\psi_2\rangle\over \sqrt 2}+ {|\psi_3\rangle\over 2}\right)\nonumber\\
\langle \rightarrow|\psi(t)\rangle &=& {1\over \sqrt 3}\op{U}(t)\left({|\psi_1\rangle\over \sqrt 2}-
{|\psi_3\rangle\over \sqrt 2}\right)
\end{eqnarray}
The state $\langle \uparrow|\psi(t)\rangle$, for example, represents the
state of the particle hitting the screen, provided that the pseudo-spin
$S_x$ has been found in the state $|\uparrow\rangle$.
The same can be represented in the position basis as follows
\begin{eqnarray}
\psi_{\uparrow}(x,t)\rangle &=& {1\over \sqrt 3}\left({\psi_1(x,t)\over 2}
+{\psi_2(x,t)\over \sqrt 2}+ {\psi_3(x,t)\over 2}\right)\nonumber\\
\psi_{\downarrow}(x,t)\rangle &=& {1\over \sqrt 3}\left({\psi_1(x,t)\over 2}
-{\psi_2(x,t)\over \sqrt 2}+ {\psi_3(x,t)\over 2}\right)\nonumber\\
\psi_{\rightarrow}(x,t)\rangle &=& {1\over \sqrt 3}\left({\psi_1(x,t)\over \sqrt 2}-
{\psi_3(x,t)\over \sqrt 2}\right)
\end{eqnarray}
Since $\psi_1(x,t), \psi_2(x,t), \psi_3(x,t)$ have already been worked out
in the preceding calculation, it is straightforward to show that
\begin{eqnarray}
|\psi_{\uparrow}(x,t)|^2 &=& {|C_t|^2\over 3}e^{-2x^2/\Omega}\left(
{1\over 2} + {1\over 2}\cosh(4xd/\Omega)\right.\nonumber\\
&&\left.+ \sqrt{2}\cosh(2xd/\Omega)\cos(2xd/a)  \right.\nonumber\\
&&\left.+ {1\over 2}\cos(4xd/a) \right)\nonumber\\
|\psi_{\downarrow}(x,t)|^2 &=& {|C_t|^2\over 3}e^{-2x^2/\Omega}\left(
{1\over 2} + {1\over 2}\cosh(4xd/\Omega)\right.\nonumber\\
&&\left.- \sqrt{2}\cosh(2xd/\Omega)\cos(2xd/a)  \right.\nonumber\\
&&\left.+ {1\over 2}\cos(4xd/a) \right)\nonumber\\
|\psi_{\rightarrow}(x,t)|^2 &=& {|C_t|^2\over 3}e^{-2x^2/\Omega}\left(
1 - 
\cos(4xd/a) \right)
\label{sxpt}
\end{eqnarray}
Clearly, $|\psi_{\uparrow}(x,t)|^2$ in the above expression does represent
a 3-slit interference. Thus we see that by detecting the particles in
coincidence with the pseudo-spin state $|\uparrow\rangle$ erases the
which-path information and brings back the interference. Quantum erasing is
possible in 3-slit interference as well.
A careful look reveals that the interference given by $|\psi_{\uparrow}(x,t)|^2$
is different from the true 3-slit interference given by (\ref{purepta}), as
can also be seen by plotting the two (see Fig. \ref{sxplot}).

The question
that arises now is, what basis should one measure the pseudo-spin in, so
that detecting particles on the screen, in coincidence with one of the
states of the basis, would yield true 3-slit interference.
Let us assume that the basis states are $|\alpha\rangle, |\beta\rangle,
|\gamma\rangle$, and that coincident detection of particles with
$|\alpha\rangle$ yields true {\em unshifted} 3-slit interference. This can happen only
if 
\begin{equation}
\langle\alpha|+\rangle = \langle\alpha|0\rangle =\langle\alpha|-\rangle \equiv c.
\end{equation}
By normalization, $|c|=1/\sqrt{3}$, and the state $|\alpha\rangle$
can be
\begin{equation}
|\alpha\rangle = {1\over\sqrt{3}}(|+\rangle + |0\rangle + |-\rangle).
\end{equation}
That normalized states $|\beta\rangle, |\gamma\rangle$ should be chosen
such that they are unbiased for the states $|+\rangle, |0\rangle, |-\rangle$,
thus satisfying the condition
\begin{eqnarray}
|\langle\beta|+\rangle|^2 = |\langle\beta|0\rangle|^2 =|\langle\beta|-\rangle|^2 &=& {1\over 3}\nonumber\\
|\langle\gamma|+\rangle|^2 = |\langle\gamma|0\rangle|^2 =|\langle\gamma|-\rangle|^2 &=& {1\over 3}.
\end{eqnarray}
Cube-roots of unity come in handy here, and the two states can be chosen as
\begin{eqnarray}
|\beta\rangle &=& {1\over\sqrt{3}}(e^{i\pi/3}|+\rangle - |0\rangle + e^{-i\pi/3}|-\rangle) \nonumber\\
|\gamma\rangle &=& {1\over\sqrt{3}}(e^{-i\pi/3}|+\rangle - |0\rangle + e^{i\pi/3}|-\rangle) 
\end{eqnarray}
Using this, the state (\ref{ent0}) can be written as
\begin{eqnarray}
|\psi\rangle &=& {1\over 3}\left[(|\psi_1\rangle
+|\psi_2\rangle+ |\psi_3\rangle)|\alpha\rangle\right.\nonumber\\
&&\left.+ (e^{i\pi/3}|\psi_1\rangle - |\psi_2\rangle + e^{-i\pi/3}|\psi_3\rangle)|\beta\rangle\right.\nonumber\\
&&\left.+ (e^{-i\pi/3}|\psi_1\rangle - |\psi_2\rangle + e^{i\pi/3}|\psi_3\rangle)|\gamma\rangle\right].
\label{entn0}
\end{eqnarray}

Particles hitting the screen can be divided into three subensembles, depending
on the state of the pseudo-spin obtained. The states of the particle,
correlated with $|\alpha\rangle, |\beta\rangle, |\gamma\rangle$, are
$\psi_{\alpha}(x,t), \psi_{\beta}(x,t), \psi_{\gamma}(x,t)$, respectively.
They can be represented as follows
\begin{eqnarray}
\psi_{\alpha}(x,t) &=& {1\over 3}\left(\psi_1(x,t)
+\psi_2(x,t) + \psi_3(x,t)\right)\nonumber\\
\psi_{\beta}(x,t) &=& {1\over\sqrt{3}}\left(e^{i\pi\over 3}\psi_1(x,t)
- \psi_2(x,t)+e^{{-i\pi\over 3}}\psi_3(x,t)\right ) \nonumber\\
\psi_{\gamma}(x,t) &=& {1\over\sqrt{3}}\left(e^{-i\pi\over 3}\psi_1(x,t)
- \psi_2(x,t)+e^{{i\pi\over 3}}\psi_3(x,t)\right ) \nonumber\\
\end{eqnarray}

The three complementary interference patterns can then be shown to have the
form
\begin{eqnarray}
|\psi_{\alpha}(x,t)|^2 &=& {|C_t|^2\over 3}e^{-2x^2/\Omega}{1\over 3}\left[
1 + 2\cosh(4xd/\Omega)\right.\nonumber\\
&&\left.+ 4\cosh(2xd/\Omega)\cos(2xd/a)  \right.\nonumber\\
&&\left.+ 2\cos(4xd/a) \right]\nonumber\\
|\psi_{\beta}(x,t)|^2 &=& {|C_t|^2\over 3}e^{-2x^2/\Omega}{1\over 3}\left[
1 + 2\cosh(4xd/\Omega)\right.\nonumber\\
&&\left.+ 4\cosh(2xd/\Omega)\cos(2xd/a+\pi/3)  \right.\nonumber\\
&&\left.+ 2\cos(4xd/a+2\pi/3) \right]\nonumber\\
|\psi_{\gamma}(x,t)|^2 &=& {|C_t|^2\over 3}e^{-2x^2/\Omega}{1\over 3}\left[
1 + 2\cosh(4xd/\Omega)\right.\nonumber\\
&&\left.+ 4\cosh(2xd/\Omega)\cos(2xd/a-\pi/3)  \right.\nonumber\\
&&\left.+ 2\cos(4xd/a-2\pi/3) \right].
\label{snpt}
\end{eqnarray}
Notice that all the three cases in (\ref{snpt}) produce true, but mutually
shifted, 3-slit interference patterns.
It implies that detecting particles in coincidence
with the state $|\alpha\rangle$, $|\beta\rangle$ or $|\gamma\rangle$ constitutes quantum eraser for the 3-slit
interference experiment. The interference, which was lost because of the
path-detector states carrying which-way information, is recovered after
the path information is erased by reading the path-detector in one of the
states $|\alpha\rangle,|\beta\rangle,|\gamma\rangle$.

\begin{figure}
\centerline{\resizebox{10cm}{!}{\includegraphics{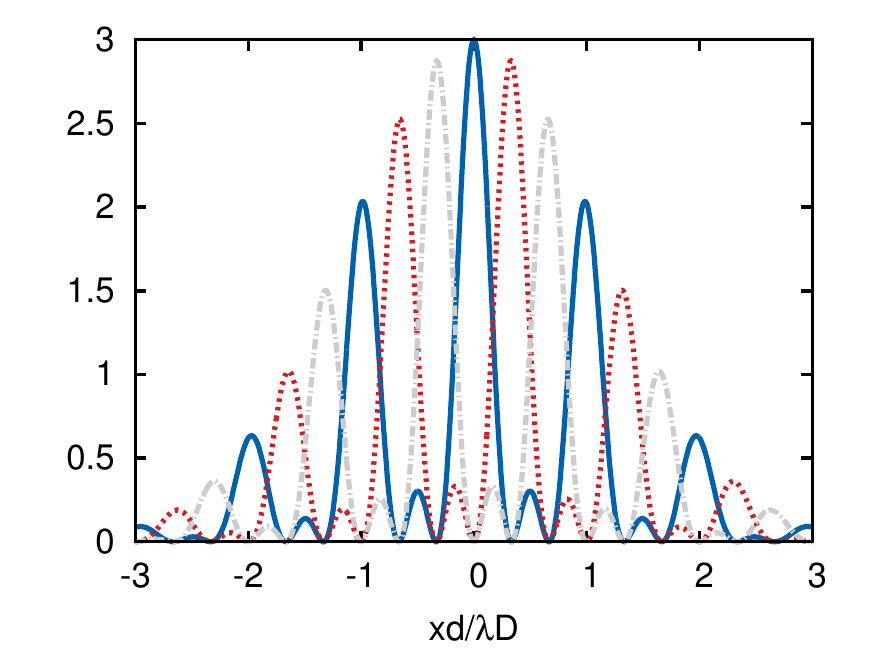}}}
\caption{Recovered interference pattern (unnormalized), given by $|\psi_{\alpha}(x,t)|^2$
(solid line), that given by $|\psi_{\beta}(x,t)|^2$
(dashed line) and that given by $|\psi_{\gamma}(x,t)|^2$
(dot-dashed line). }
\label{snplot}
\end{figure}

\section{Discussion}

In the preceding analysis, we have shown that the concept of quantum
eraser can be extended to the case of 3-slit interference. If the which-path
information, of the interfering particle, is stored in a quantum path-detector,
the interference is lost. By erasing
the path information stored in the quantum path-detector, the lost 
interference can be made to come back.

In a typical implementation of quantum eraser for a 2-slit interference
experiment, the particles are detected in coincidence with two orthogonal
states of the path-detector. For example, in a particular implementation
with photons passing through a double-slit, they were detected in coincidence
with two orthogonal polarization states of the photon \cite{eraser10}.
Both these states lead to recovery of true 2-slit interference. We have
shown that in the case of 3-slit interference too, three complementary
interference patterns can be recovered.

A quantum eraser for 3-slit interference will definitely be harder to
implement, as compared with 2-slit interference. However, we feel that
an earlier proposal for quantum eraser for 2-slit interference using a
modified Stern-Gerlach setup \cite{zinitq}, may be a good candidate for
extending to 3-slit interference.

\section*{Acknowledgement}
Naveed is thankful to the Centre for Theoretical Physics, Jamia Millia
Islamia, New Delhi, for providing its facilities during the course of this
work.


\begin{thebibliography}{0}

\bibitem{young} T Young, 
{\em Philosophical Transactions of the Royal Society of London} {\bf 92} 12-48 (1802)

\bibitem{jonsson} C J\"onsson, {\em Am. J. Phys.} {\bf 42} 4-11 (1974) 

\bibitem{tonomura} A Tonomura, J Endo, T Matsuda, T Kawasaki and H Ezawa,
{\em Am. J. Phys.} {\bf 57} 117-120 (1989)

\bibitem{bohr} N Bohr, {\em Nature} {\bf 121}, 580-591 (1928)

\bibitem{tqeinstein} T Qureshi, R Vathsan, {\em Quanta} {\bf 2}, 58-65 (2013)

\bibitem{dsen} D Sen, {\em Pramana - J. Phys.} {\bf 72}, 765-775 (2009)

\bibitem{wootters} W K Wootters and W H Zurek,
{\em Phys. Rev. D} {\bf 19}, 473 (1979)

\bibitem{greenberger} D M Greenberger, A Yasin,
{\em Phys. Lett. A} {\bf 128}, 391 (1988)

\bibitem{englert} B-G Englert, {\em Phys. Rev. Lett.} {\bf 77}, 2154 (1996)

\bibitem{jaynes} E Jaynes, in {\em Foundations of Radiation Theory and
Quantum Electronics}, ed. A O Barut (Plenum, New York 1980), pp. 37

\bibitem{scully} M O Scully, Kai Dr\"uhl, {\em Phys. Rev. A} {\bf 25}, 2208 (1981)

\bibitem{eraser1} P G Kwiat, A M Steinberg, R Y Chiao,
 {\em Phys. Rev. A} {\bf 45}, 7729 (1992)
\bibitem{eraser2} T J Herzog, P G Kwiat, H Weinfurter, A Zeilinger,
{\em Phys. Rev. Lett.} {\bf 75}, 3034 (1995)
\bibitem{eraser3} S Durr, T Nonn, G Rempe, {\em Nature} {\bf 395}, 33 (1998)
\bibitem{eraser4} B Dopfer, Ph.D. thesis, Universtat Innsbruck, (1998, unpublished)
\bibitem{eraser5} P D D Schwindt, P G Kwiat, B-G  Englert, 
{\em Phys. Rev. A} {\bf 60}, 4285 (1999)
\bibitem{eraser6} Y H Kim, R Yu, S P Kulik, Y Shih, M O Scully,
{\em Phys.  Rev. Lett.} {\bf 84}, 1 (2000)
\bibitem{eraser7} T Tsegaye, G Bjork, M Atature, A V Sergienko, B E A 
Saleh, M C Teich, {\em Phys. Rev. A} {\bf 62}, 032106 (2000)
\bibitem{eraser8} P Bertet, S Osnaghi, A Rauschenbeutel, G Nogues, A
Auffeves, M Brune, J M Raimond, S Haroche, {\em Nature} {\bf 411}, 166 (2001)
\bibitem{eraser9} A Trifonov, G Bjork, J Soderholm, T Tsegaye, {\em Eur.
Phys. J. D} {\bf 18}, 251 (2002)
\bibitem{eraser10} S P Walborn, M O Terra Cunha, S Padua, C H Monken,
{\em Phys. Rev. A} {\bf 65}, 033818 (2002)
\bibitem{eraser11} U L Andersen, O Glockl, S Lorenz, G Leuchs, R Filip,
{\em Phys. Rev. Lett.} {\bf 93}, 100403 (2004)
\bibitem{eraser12} G Scarcelli, Y Zhou , Y Shih, {\em Eur. Phys. J. D} {\bf 44}, 167-173 (2007)

\bibitem{jaeger} G Jaeger, A Shimony, L Vaidman, {\em Phys. Rev. A} {\bf 51}, 54 (1995)

\bibitem{durr} S D\"{u}rr, {\em Phys. Rev. A} {\bf 64}, 042113 (2001)

\bibitem{bimonte} G Bimonte, R Musto, {\em Phys. Rev. A} {\bf 67}, 066101
 (2003)

\bibitem{englertmb} B-G Englert, {\em Int. J. Quantum Inform.} {\bf 6}, 129 (2008)

\bibitem{3slit}  M A Siddiqui, T Qureshi, {\em Prog. Theor. Exp. Phys.} {\bf 2015}, 083A02 (2015)

\bibitem{g3slit}  M A Siddiqui, {\em Int. J. Quant. Inf.} {\bf 13}, 1550022 (2015)

\bibitem{coherence}  M N Bera, T Qureshi, M A Siddiqui, A K Pati, 
{\em Phys. Rev. A} {\bf 92}, 012118 (2015)

\bibitem{bagan}  E Bagan, J A Bergou, S S Cottrell, M Hillery, {\em Phys. Rev. Lett.} {\bf 116}, 160406 (2016)

\bibitem{nslit}  T Qureshi, M A Siddiqui, 
{\em Ann. Phys.} {\bf 385}, 598-604 (2017) 

\bibitem{urbasi} U Sinha, C Couteau, T Jennewein, R Laflamme, G Weihs,
{\em Science} {\bf 329(5990)}, 418-421 (2010)

\bibitem{neumann} J von Neumann, {\em Mathematical Foundations of Quantum Mechanics}
(Princeton University Press, 1955)

\bibitem{kaur} M Kaur, B Arora, M Mian, {\em Pramana - J. Phys.} {\bf 86}, 31-48 (2016)

\bibitem{beau} M Beau, T C Dorlas, {\em Int. J. Theor. Phys.} {\bf 54} 1882 (2015)

\bibitem{zinitq}  T Qureshi, Z Rahman, {\em Prog. Theor. Phys.} {\bf 127}, 71 (2012)

\end{thebibliography}
\end{document}